# Crisis` Heritage Management
## New Business Opportunities Out of the Financial Collapse

*"He Who Controls the Past, Controls the Future." (G. Orwell)*

Hristian Daskalov

Brain Workshop Institute,

Policy Expert

*e-mail: hdaskalov@cobdenpartners.co.uk*

**Abstract**: This paper intends to present the opportunities emerging for the national economy, out of the financial crisis. In particular the management of those, which arise from the commercial real estate owned property sector, defined by the author as a „*crisis' heritage management*". On one hand, as real estate property prices are subject of wide fluctuations, the longer possession of such assets can seriously impact the financial condition of the already shattered financial institutions, but on the on other, with the help of professional and proactive management and the right kind of attitude by all the stakeholders, the „*heritage*", left out of the financial collapse, can not only help stabilize the system - bringing liquidity into it, but can also support its healthy corporate governance in the long-term. The properties themselves (*business buildings, warehouses, retail- and office spaces*), being an object of optimization of maintenance costs, reengineering, intensive marketing, as a result of the crisis, can serve as a solid base for number of new and profitable business and investment opportunities, described in the present article, as a proof of the healing effect of the financial crisis and the „second chance" it gives.

# 1. Introduction

In the beginning of every analysis of an economic problem two common understandings, accepted as „laws" in economy, should always be reminded. To each economist and his theories opposes another one, equally recognized and honoured, but antagonistic in his views on the problem. And the second understanding – they both, in most of the cases, turn out to be wrong. Though those are just humorous statements not claiming for accuracy, it cannot be left unnoticed that economists and practitioners, differing in their fundamental views on subjects such as financial stability and economic growth, do not attempt to work together on more practical issues where a common ground can be reached, mainly because of the differences in their basic economic understandings. Therefore this paper is dedicated neither on the causes behind the financial crash, nor on the after-effects alone, as those are topics *„too big to agree"*, but instead is concerning the „leftovers" that can be either easily labelled as bad assets which we should all get read off and put aside of the recovery path, or to be looked at as a beginning of a whole new value chain that is to be a subject of special managerial, entrepreneurial and administrative efforts.

Quite too often when attacking issues such as the economic recovery and the recovery of the financial sector in particular we turn out to be in situations where the *how*-part, instead of the *what*-part, is not just the primary one, but the sole subject of experts' guidance. This distracts stakeholders' attention from the core of the problem itself, turning it to the broader discussion on the global layout of the problem's resolution. In other words, experts argue how to cope with something which they haven't closely monitored, neither have quite well comprehended as in the case of commercial real estate owned (OREO) property[1]. For example, states desperately want to restore confidence in the financial system, but do not improve its management. Instead they focus on the „*how*"-part and not surprisingly give the wrong calls. They just insist again and again on restoring the status quo, believing it will bring back *„the old good days"* along with it. But should those old days actually be labelled as „good"? Why don't we look on what we have left not as a burden, but more like a heritage and direct our efforts towards the better management of the system's heritage - what is it, what is its value, what it costs to maintain and improve, what kind of systemic weaknesses unveils and what kind of opportunities does it hold? Answers on those questions are being seek in the present article.

---

[1] According to Roark, W. (2006), Concise Encyclopedia of Real Estate Business Terms, real estate owned or OREO is a class of property owned by a lender - bank, government agency or government loan insurer. The terms "REO" and "OREO" ("other real estate owned") are often used interchangeably

## 2. Financial sector's creative destruction

First of all, financial crises are all around us (*historically*), and whatever we do, overall, we cannot prevent them from accruing if to follow sound and reasonable principles in economy. Especially banking crises. 50 major banking crashes across Europe and the US are described dating back to 1618 (Kindleberger). The World Bank on the other hand have identified more than 130 countries with a banking crisis since the 1980s (Segura). If we look through one of the best books on the present crisis - *„This Time Is Different. Eight Centuries of Financial Folly"* we shall see that throughout 800 hundred years of history there is an extraordinary range of financial crises with global outreach. Typically, countries with sound bank regulations and supervision are capable of handling with such crises, but on the other hand, it's obviously not typical for them to be capable of reaching a consistency in their level of policy soundness – on strategic and operational bases. Plus both regulatory and supervisory policies are far from what can be described as *„the one and only"* adequate predictor, given that regulation is so volatile in terms of content, quality and direction in which is leading the system to. Not to forget that regulations often are the greatest and most powerful cause preventing us from spotting a failure.

One is for sure – the impact on the real sector after any given banking crisis is great, starting from the overall economic growth which is affected[2]. We won't be looking in details on the impact on monetary policy which can be large, in an enormous extend with exchange rates' levels, inflation, etc; the impact on fiscal policy, governmental budgets, etc. But is this impact all negative? Again, without trying to theorize to a great extent, we cannot miss to translate Schumpeter's theory on creative destruction for the purposes of the present topic. But we will neither turn it into a defence on the theory itself, as the functional importance of a failure in the economic system is already a well defended thesis. And practically a proven one. In the end, those who do not provide satisfactory goods and services, should and will eventually fail. The question is will they bring the system down with them if not allowed to (by *the exact same system itself*), and if so, is that unjust or just what is needed to happen? Without an exit-alternative, can we even discuss for basic signs of capitalism. Here is where the difference is. When speaking of banking crises at present (*the ones that are not*

---

[2] The international monetary fund estimates fiscal costs can go above 20% of GDP in some of cases - Argentina and Mexico in the 80s and 90s are only a few examples.

*speculative in their nature, with short-term liquidity problems)*, it is not about the „*decentralized system of choices*" that leads to a failure. Rather than that, it is the „*centralized system of mismanagement*" behind, which leads to far more destructive consequences for the economy. This may be a significant difference, but one that does not change the creative character of the destruction itself. And the non-exit alternative[3], most commonly referred to as „*solidarity*" in Europe, is just a sign of such crisis mismanagement, if the occurrence of the crisis itself we can refer to as a form of operational mismanagement. The other approach on the subject will further deepen the distance between markets and banks – in terms of sound managerial principles' implementation and use. It is the zombie-economy approach in which insolvent and in fact, bankrupt companies, banks, governments continue to operate despite their "mortality", as a result of which this becomes a heavy burden for the "healthy" segments of the economy[4].

To prove the theoretical statements claimed so far, and eventually reach the essentials of the theme, we should look at the core of the crisis itself, for reasons no different than to establish an interface, connecting the global financial risks, challenges and business opportunities that occur out of it, with the situation in Bulgaria in particular, which is a case-study ground of a kind, though we all know the birthplace of the crisis – US' subprime loans market. Note that we refer to the „*market*", not to the financial instruments themselves, as the trade with bad mortgage loans, and particularly their role as collateral for various types of securities, led to the expansion of what followed on a global scale. Prior to that, dating back to 2001, was the main „*suicidal incentive*" that triggered this chain of events - the low interest rates in the US which led to increase in the number of mortgage loans and a furious rise in prices of real estate property, undervaluing it's quality *(which we will later refer back to as a core principal of the improvement processes)*. The subsequent transfer and re-transfer of the risk by banks to investors through "securitization" in search of liquidity does not need further explanations for its inconsistency to be described. The links between the events on the other hand such as the transformation into a public finance crisis, do require further comments on them so that the systematic character of the crisis can be observed and analyzed for the best solutions to be drawled out of it.

---

[3] No less important than the freedom to enter a market is the freedom to leave. (Frank 2009: 214)

[4] Takeo Hoshi and Anil K. Kashyap, "Solutions to Japan's Banking Problems: What Might Work and What Definitely Will Fail" in „Reviving Japan's Economy: Problems and Prescriptions", ed. Takatoshi Ito, Hugh Patrick and David E. Weinstein (Cambridge, MA: The MIT Press), 2005;

If we agree that the crisis of trust at the interbank money market, gave the spark for the banking crisis, followed by recession, then we can connect it[5] to the present day's public finance crisis in Europe, accompanied by public debt, budget deficits, private debt and recession of second phase that would eventually lead to stage three of middle and long-term deep recession across European economy, followed by social unrests (*already happening*) and unclear and uncertain future. This „stage three" that is yet to come is to be tackled most effectively *by* focusing on the legacy, on the heritage left after the first two waves of the crisis, instead of denying its existence and the opportunities that is creates. So if we are to exit successfully out of the situation we are presently in, it is not about inflating the financial balloon again, this time with governmental bonds, what is public officials' and bankers' main activity in the last couple of years, but to focus on what we already posses as a heritage of the crisis and manage it in a profitable way so we can eventually win back the future for entrepreneurial capitalism where in order to grow, you need production and movement of real goods, not production and movement of false accounts and policies that depreciate the value of real assets management or lead to self-devaluation of the assets possessed. The poor and systematic lack of liquidity that EU banking system is associated with, can be managed not by financial, political and administrative engineering, inevitably followed by „*side effects*" such as boom of speculation, moral hazards, inflation, etc., but by understanding the importance of intelligent and proactive real assets' management. A lesson that should be learned also by the national governments, concerning their own properties and resources, which represent a far more wise collateral than the „*printing press*" resolution, especially as over the past few years, an alternate REIT[6] segment has emerged, comprised of owners of income-producing real estate such as timber, data centres, document storage facilities, cell towers, even prisons and billboards, among which governments should consider their place[7].

---

[5] The author points out a clear interrelationship between the situation in the US and Europe in terms of economic catalysis. Though the present burst in the EU began formally in 2009 with the discovery of Greece's faked accounting *(the fiscal deficit turned out to double itself form 6% - officially, to 13% of GDP in reality)*, in addition to the high levels of debt, given the unwise levels of "social" benefits throughout the EU and the unwise borrowing and lending – both public and private (not without the help of the ECB policies), when the financial crisis from the US reduced fiscal budget revenues, all the EU countries suffered very high fiscal deficits. Those deficits led to difficulties in affording „big-spender" policies, but yet again they were financed with new debt, and eventually things got even worse.

[6] A real estate investment trust (REIT), is a company that owns income producing real estate such as offices, hotels, healthcare facilities, apartments, shopping centers, etc. It operates under different regimes in the different countries since it first originated in 1960 in the US. Its original idea was to make investments in large scale income producing real estate accessible to smaller investors. Generally REITs pay little or no corporate income taxes because they are able to deduct dividends they pay from their taxable earnings.

[7] Deloitte Centre for Financial Services (2013), Non-traditional commercial Real Estate: Capitalizing on the REIT opportunity.

# 3. Institutional Overview of the Crisis' Heritage Management

## 3.1. Macro Aspects

What makes a banking problem systemic and turns it into crisis? - The scale – whether is related to just a bank or the entire system. There is one more circumstance. Banks may have short-term liquidity problems, or can have no problems to meet their obligations what so ever, but if in the long run banks equity is eroded by high number of non-performing loans, then this is a sure sign for trouble ahead and measures should be taken. On the other hand, we've had a chance to witness a number of state interventional measures, none of which acceptable by the taxpayers anymore, including acquisition of distressed assets through different forms of state structures with low probability of market success and high risk of growing loses with the time passing by – the example with the performance of the National Asset Management Agency (NAMA) in Ireland speaks for itself[8]. Not to mention the alternative for long term intervention into the bank's financing and management. So the right set of measures should be one with strategic horizons, not the short-sighted approach as short-sightedness too often may lead to governments writing checks with numbers „*picked out of somebody's ...*" as in the case of Anglo-Irish Bank[9] where the bottom-line was that the bank's management decided to game the Central Bank and the regulators knowing that once the State (of Ireland) began the flow of money, it would be unable to stop: *"If they (Central Bank) saw the enormity of it up front, they might decide they have a choice. You know what I mean? They might say the cost to the taxpayer is too high . . . if it doesn't look too big at the outset... if it looks big, big enough to be important, but not too big that it kind of spoils everything, then, then I think you have a chance. So I think it can creep up... [once] they have skin in the game."*[10]

In order to recover maximum value at minimum cost – whether we address the issue of non-performing loans or real estate assets management, banks, banking authorities, even governments should strictly keep to the market-based solutions and rely on their personal liability through sound ex-ante assessments of their possessions and proper surveillance including asset reviews, etc. And also implement thorough strategic analysis of options before any proactive and reactive measures are taken, depending on the circumstances which we cannot ignore. This opens a tempting niche for new high value businesses to occur because in the current market environment, the demand for smart governance of the financial system is huge and companies that offer their expert services on a non-prescriptive, case-by-case basis, are to win much out of the collapse.

---

[8] NAMA, the „*bad bank*" set up in 2009 to clear toxic property loans from Irish bank balance sheets (*though eventually the country was forced to accept a €67.5bn bailout from the EU and International Monetary Fund*), didn't sell a big tranche of commercial property loans until May 2013 when it successfully sold loans linked to Irish property and land with a par value of €800m to a joint venture company in which retained a 20 per cent stake. The loans were sold for about €200m, which amounts to a 75 per cent haircut on the original loan value. The deal was announced as a „*huge success*" that would kick-start a much wider sale (*NAMA's overall portfolio is about €8bn.*)

[9] More on the Anglo-Irish bailout and the journalistic discoveries behind it at: http://www.independent.ie/business/irish/inside-anglo-the-secret-recordings-29366837.html

[10] Anglo-Irish's senior manager John Bowe speaks about how the State had been asked for €7bn to bail out Anglo – but Anglo's negotiators knew all along this was not enough to save the bank. Anglo itself was within days of complete meltdown – and in the years ahead would eat up €30bn of taxpayer money.

On a European level things look exactly the same – with a lot to improve on an organizational level. Most of the initiatives on the EU, concerning banking system stability, are exactly in the wrong direction and have nothing to do with what's in the core of the problem. What kind of idea is the establishment of „*Common Eurozone Bonds*" for example in times when the ECB is suspending national government bonds as collateral in the Eurosystem monetary policy operations, following cuts in national credit ratings to selective defaults[11]? To cure a system, suffering from deficit of personal liability with the same, that caused the problem, but in bigger quantities and to continue and further develop the Ponzi scheme in which failed banks and supporting failed states (*and vise-versa*) with even bigger flow of printed money, pumped into the system by the ECB (*the so-called zombie-economy case*), is a recipe for total disaster. Equally irresponsible is to continue lending money to banks that nobody wants to have nothing in common with and no wonder as they have worked under the principle of „*privatization of profits - nationalization of losses"* or to transfer the responsibility to supranational level as the project for a new EU agency to wind down troubled banks would do with all the moral hazards ahead[12].

Instead, the EU officials should refocus their attention back to the market basis of both the threats and the opportunities unveiling in front of the system. This is, of course, an invocation in a second best situation where they (*the EU institutions*) are here to stay playing an active role in the overall system governance. But this role should more of an analytical character, providing a comprehensive process framework, while in the same time staying on the sidelines of the processes. Isn't there a need for joint public-private efforts to comprehend better the changes in Europe's commercial real estate market, for example; to assess the property debt developments; to suggest models for better public sector real estate management, having in mind that EU is a common market[13]? Europe houses around 40% of the worlds' commercial real estate but only 14% of the world's listed property market, according to the European Public Real Estate Association (EPRA). Hence, European policy makers have a huge opportunity to improve the contribution that the real estate sector makes in, driving the economy towards smart, sustainable and inclusive growth, by tailoring such policies that encourage the expansion of those companies[14], which help improve transparency and accountability in property market data[15], and by their nature offer a highly-liquid real estate investment opportunity for individual investors[16] - with diversification and lower levels of risk[17] and higher income and capital returns, making OREO properties, included in the portfolio of such companies, once again attractive and profitable, thus contributing to the recapitalization and greater levels of liquidity in the financial sector (*and in the otherwise illiquid property market*) and contributing to the real economy through the services provided (*in delivering the built environment and in creating those accessible investment opportunities mentioned above*). Unfortunately, EU policy makers have in many times an agenda, which differs from basic markets logic and follows its internal, best described by the words of Ronald Reagan – „*If it moves, tax it. If it keeps moving, regulate it. And if it stops moving, subsidize it.*"

---

[11] Cypress' case. More on the subject at: http://www.bloomberg.com/news/2013-06-29/ecb-suspends-cyprus-government-bonds-as-collateral.html

[12] The body as projected by the EC would form part of a banking union, designed to underpin confidence in the euro zone and end the chaotic handling of cross-border bank collapses, performed so far by the same engineers of the project – the EC and the ECB, eventually with the same amount of success.

One is for sure – the chronic systemic malfunction – the one we presently observe, due to the constantly devaluating assets and the losses accumulated, lead to an overall loss in financial sector's efficiency around Europe. If we add to that an increase in the number of bankruptcies among private entities, as well as cases of actual insolvency, slow bankruptcy procedures and other business related administrative barriers and obstacles, this inevitably leads to an even bigger increase in the overall level of indebtedness on a national and EU level which serves good to nobody. Hence it's in the interest of all stakeholders to take their share of responsibility and turn, through active management approach, this great weakness and threat to the long-term economic recovery that we speak of, into an opportunity to strengthen the financial and public sector governance and gain from the new value-added business opportunities. This is a far better solution for „*mobilization of resources*" than resorting solely on publicly funded bailouts, which only worsen the situation, as a consequence of the lack of strategic efforts. Because the resources are there, they don't need to be created out of thin air. Constructing comprehensive real estate policy a measure to protect the liquidity of the national banking systems and the one of the individual banks alone, but also ensures their ability to continue to extend credit to the economy *(improves resource allocation);* strengthens corporate governance and accountability *(this also refers in even greater extends to public asset valuation and management)* and strengthens competition in the financial world, based on solid market principles. Accountability, credit quality, and time are at stake.

---

[13] Concerning the new risks, unique to the rental of REO residential properties, with regards to one broader view on the global practice, a European-wide comparative report with best-practices and benchmarks can be initiated by the EC, dealing with issues such as re-rental of previously occupied properties; liability risk arising from rental activities, along with the use and management of liability insurance or other approaches to mitigate that liability and risk; legal requirements arising from the potential need to take action against tenants for rent delinquency, potentially including eviction. All of which are a subject of systemic legislative review that can be of use for the system's overall improvement. Source: *Federal Reserve Policy Statement on Rental of Residential Other Real Estate Owned Properties, April 2012*

[14] Stock Exchange Listed Property Companies. Building a Stronger Europe, European Public Real Estate Association, 2013

[15] Findings of the Jones Lang LaSalle's 2012 Global Real Estate Transparency Index show that a healthy and sizeable property sector improves on an national level the overall transparency of a country's real estate market, as a result of which the state develops an advantage in attracting international capital and in supporting domestic investor allocations to real estate. More at: http://www.joneslanglasalle.com/GRETI/en-gb/Pages/Global-Transparency-Index-KeyFindings-2012.aspx

[16] Although objects of real estate investments are long-term assets, the listed market in which REITs function provide individualized time-span investment opportunities by buying and selling shares at any time. Liquidity itself improves the flow of information, allowing good management decisions to be rewarded and bad ones - punished by investors quite timely, with a strong and important corrective-behavioural effect on companies' management.

[17] According to the EPRA's „Stock Exchange Listed Property Companies. Building a Stronger Europe" publication: „*The listed property sector reduces information-based contagion which is a key contributor towards systemic risk by reducing the likelihood of opaque market bubbles and subsequent market shocks.*"

### 3.2. Considerations on a National Level

What we need above all for a profitable resolution is to focus the main efforts on a national level, not on EU level, as the country-individual stakeholders are capable of overcoming this problem with non-prescriptive, case-by-case solutions, without of course any contradiction with the previously stated recommendations and having in mind the need for pan-European overview, assessment and best-practices recognition. As with every other problem of the national economy, this also demands specific methods and techniques for problems-solving inside the national finance and business environment and tailed policies following some basic principles such as the liberalization in the legislative system. In particular, a need for liberalization, concerning the possession and management of commercial real estate property by lenders, allowing the financial sector to gain liquidity after the peak of foreclosures in the present times of low real estate prices and low demand. Liberalization, instead of regulations, can stream the entrepreneurial initiative in the prosperous sector, connected with the management of such assets and the investment opportunities linked with them.

Adopting sector-specific reporting practices is also a good branch initiative for sector's administrative and regulative decentralization form the supranational bureaucratic guidance, which have already taken place among the stock exchange listed property companies, for example, including REITs *( which acquired and are presently in possession of vast property portfolios, including residential, commercial, retail and industrial real estates, previously foreclosed)*, responding to the demands of investors which insist as a result of the crisis for even more detailed, consistent and sector-to-sector specific information to enable them to make more informed decisions, with regards to their investments, as many finance and investment experts are sceptical that the International Financial Reporting Standards (IFRS) alone, can provide the level of assurance on the quality of information in possession of shareholders, creditors and other stakeholders in the business, that they demand for[18]. In order to demonstrate the differences, a detailed look at the key performance indicators (KPIs) is needed. Under the IFRS, investment property is to be recorded at fair value or cost (l*eading to less depreciation*). Under the sector-specific reporting practices it can be only recorded at fair value (*e.g. price to exchange property*). Plus net asset value (NAV) and earnings per share (EPS) are tailored to investors under EPRA's BPRs and other performance indicators such as Net Initial Yield and Vacancy Rate are also tailored to investors.

The light in the present article is on commercial property as it is an income-producing property that both financial and public sector institutions are presently in possession of in large amounts[19] (*if we use a broader definition adding to the group property assets such as hospitals, institutional office-buildings and other public sector property plus industry and retail real estate*). But residential property and its active management is also a significant challenge that and is not excluded from the overall assessments made in the article. For example, in its policy statement on rental of residential other real estate owned properties[20] from April 2012 in the light of the large volume of distressed residential properties and the indications of higher demand for rental housing in many markets, the Federal Reserve made a point, that financial institutions can make greater use of rental activities in their disposition strategies (*whether acquired through foreclosure or voluntary surrender of the property by a seriously delinquent borrower*), reminding that the Fed's regulations and policies permit the rental of residential other real estate owned (OREO) properties to third-party tenants as part of an orderly disposition strategy within statutory and regulatory limits *(such as the time of possession limit)*. Rental property is not a passive investment, it's a business. You are in charge of buying of finding tenants, collecting rents, ensuring that maintenance and renovations are done, etc.; hence this creates opportunities for outsource businesses to manage some or all of these tasks while the financing institution remains responsible for ensuring that they get the job done. Commercial property on the other hand is a more interesting subject of analysis because it has a longer value-chain and if successfully turned into use for the real economy can have a huge impact on bringing it to the path towards growth *(though residential property also posses an entrepreneurial potential)*. It creates around itself a whole new vivid business environment, consisting of consultants, special servicers, other third-party vendors whom primary business is to innovate - by introducing new ways to put into use the other real estate property and create markets for it.

---

[18] Gordon Kerr, the founder of Cobden Partners and the author of "The Law of Opposites: Illusory Profits in the Financial Sector", is among the experts, sceptical that the IFRS are capable of providing prudent accounting. Some even argue that those accounting standards are "destabilizing banks" and "damaging national economies". http://www.bloomberg.com/news/2013-03-11/europe-s-accounting-rules-are-destroying-its-banks.html
Therefore, according to the author of the present article, it is a wise solution to have a sector-specific reporting practices, enhancing the quality of IFRS information, such as the so called „Best Practices Recommendations" (BPR), imposed by the European Public Real Estate Association with adoption levels at 85% of the European listed property companies (according to EPRA, http://www.epra.com/).

[19] The capability to generate a strong, consistent income stream makes commercial real estate also more attractive for investment even in a slow-growth environment and in a situation, as the present one we have today around Europe, where the continuingly slow interest rates have limited the yields and appeal of fixed income investments.

[20] The term "residential properties" in the policy statement encompasses all one-to-four family properties and does not include multi-family residential or commercial properties.

Perhaps the highest form of innovation by an entrepreneur is the creation of new forms of organization, as Schumpeter claims, thus causing continuous progress and improvement for everyone. This process of reorganization of property assets can therefore be labelled as essentially useful for the national economy; its facilitation – as needed. The latter can be used by some to justify a limited governmental role in the process, but as mentioned previously – only at the sideline. Public sector officials should not be tempted to actively intervene as they intend to do in Ireland where a big-scale social housing project is creeping around the corner with great number of residential properties, controlled by NAMA's debtors, as if the housing policy and backing it up with the support of financial engineering was not in the basics of the housing bubble and the following collapse in the US[21]. Instead they have to consider bringing state's own real estate property to the minimum and what's left - to assess and capitalize on the markets though special entities such as a sovereign wealth fund (SWF) of a kind or through public sector real estate investment trust or another vehicle, comprising government-owned asset pools. Sooner or later such steps are inevitable to be taken by the nations. The question is whether they would be proactive and beneficial for the state or pro-cisis, in situations of despair, as in the case with the developments on Greece's bailout[22]. The difference between those two scenarios is in the timing and the levels of political acumen. In this line of reasoning the case of the over performing Government Properties Income Trust on the stock exchange should be analysed as a potential role model, for design of government property management model, though in the case of the trust mentioned, it is about a private fund leasing properties to the U.S. government[23]. Such bodies would have the chance to replicate the success of the standard REITs as innovators in delivering and operating the built environment, responding more actively to the needs of the citizens[24] and have a positive effect in terms of professionalism, transparency, stability, liquidity and quality of the management. Not to forget the role of such innovation in driving up the standards and efficiencies in the governance sector in a broader sense and with broader long-term implications.

---

[21] NAMA has identified over 3, 800 residential properties controlled by its debtors and receivers as being available and potentially suitable for social housing provision. The process of assessing suitability and confirming demand is on-going for part of the units. Many of those confirmed to date are already in the process of negotiation with Approved Housing Bodies (AHBs) for either acquisition or leasing. The Agency's identification of over 3,800 units represents potentially one of the single largest allocations of social housing in the history of the State. (source: http://www.nama.ie, as of 10/07/2013).

### 3.3. Bulgaria's Case

How all of this looks on Bulgarian ground in particular? Due to the relative backwardness in Bulgarian banking system, its smaller capacity and use of complex financial instruments, plus the surplus and large fiscal reserve of Bulgaria in the years prior 2008, the „*first wave*" of the financial crisis was less severe, but the overall economic effects do not differ form the situation in the EU. The strong punch on the banks in Bulgaria came directly from the construction sector, affected by the global conjuncture, which didn't allow its further boom that eventually would have led to even more serious consequences *(as the ones that*

---

[22] Back in 2011 Eurozone governments were discussing to have a noncash Greek governmental assets, including real estate, offered as collateral for a new round of rescue lending as part of the overall bailout plan. More at: http://online.wsj.com/article/SB10001424053111903461304576526613839378594.html

[23] As of March 31 2013, Government Properties owned $1.7 billion in office properties, comprising 10 million square feet in 31 states and Washington D.C. The company claims that about 75% of its rental income comes from properties leased to the U.S. government. In addition, 21% of its rental income is paid by state governments and 4% by the United Nations.The dividend at Thursday's closing share price of $23.36 translates to a 7.36% yield, well above the average REIT yield. Plus, government tenants have a historically high renewal rate with an average occupancy of 26 years. Source: http://finance.yahoo.com/news/government-reit-trust-131000281.html

[24] Recently, in March 2013, government-backed panel in Japan recommended launching healthcare REITs to help finance the construction of elderly care facilities, an investment venture by the Shinsei Bank Ltd. which would initially purchase assets worth 100 billion yen ($1 billion). A public-private partnership in this social area would be interesting to observe and replicate in Europe. Securitisation is sometimes, as in this case, the best solution for vast investments in sectors as health and social care as such projects need money from day-one but debt is paid back over a span of sometimes 15 to 20 years. Germany is presently in a need for new senior housing and assisted living facilities. Such need exists Europe-wide, and will continue to grow given the demographics, similar to the Japanese case. This represents an attractive investment opportunity for research on a joint REITopportunity, powered with undervalued (presently) REO property in countries such as Bulgaria, having the social, economic and built-environmental conditions.

*Latvians experienced)* if the time for the global and national economy to sober had delayed with a year or two. And no wonder as of about 27-28 billion FDI for the period 2005-2009 in Bulgaria, approximately 70-75% were in the sectors of finance, construction and real estate (*residential, resorts, shopping centers, etc*.)., trade, which in total number represents about 19 bln. euro, most of which is now "frozen"[25].

At present[26], the dynamics of the balance sheets in the banking system are not under the influence of fundamental changes. Borrowed funds continue to grow, while bank lending remains weak with resources, allocated mainly to low-risk investments - investments in credit institutions and securities. As a result, the liquid assets of the banking system and are at levels of about 25%, with increase in both revenues and profits. Management of problem loans and allowances for provisions on the other side continues to be a major challenge for banks for reasons already mentioned. In the third quarter of 2012 the level of restructured, under-performing, non-performing loans was approximately 25,8 per cent, well above the average levels for the countries in the region of Central Europe. Nearly 90% of households in Bulgaria live in their own homes, with the share of the mortgaged property is less than 1%, according to figures from Eurostat, on housing conditions in Europe at the end of 2011. Clearly, according to this statistics, Bulgaria is far from Ireland's residential property troubles and that is why, due to the significant differences around Europe, is not appropriate to tackle the issue of OREO management with common prescriptions.

Comparing to other countries such as Spain, where the banks are already starting to reassess their strategies to sell their shares in the joint companies for OREO property management to foreign investors, as in the case of La Caixa Bank and Servihabitat Gestion[27], and Bankia and Habitat, Bankia's property management company, in Bulgaria banks are just starting to establish their own agencies, on which the next chapter of the article is dedicated to. And it's not too late for them because the demand for retail and office space in Bulgaria right now is rising, mainly driven by international players - outsourcing and IT companies entering the market, as well as from pharmaceutical companies. More and more companies start looking for opportunities to expand its portfolio with investments in countries such as Bulgaria due to the low taxes, highly skilled workforce and undervalued properties. The market for logistic business space in the country is also growing, with a positive trend for the next 5 years, according to specialists[28].

---

[25] The numbers are approximate, from the National Investment Agency.

[26] The data is from Bulgarian National Bank report on the financial system from May 2013.

[27] It is not clear what's the value of ServiHabitat as a standalone business, but the income from rental and property management services of the group La Caixa overall in 2012 almost doubled, reaching levels of up to $ 1.6 billion.

It is up to the OREO management sector to play its cards well, improve the value of its possessions and provide a cost-effective way for financial institutions to maximize possible earnings on the foreclosed commercial real estate owned properties. Every euro spent on such efforts is euro, invested in the real economy smart and sustainable growth, from where come the mid-and-long term returns. Because as the economic conditions improve, the market value and the price of real estate owned property also increase, hence there is a direct correlation between the efforts put by the owners and the economic results achieved. Not to forget that commercial property sector is essentially important for the well-being of the national economy, being directly responsible for providing this essential need of businesses and citizens –space and infrastructure for economic and social purposes[29]. And investing in the built environment means investing in providing those needs, no matter the present ownership – whether in the hands of institutional, or individual investors *(through the real estate investment trusts, for example).*

In this line of thoughts, linking crisis' heritage management at a national level in Bulgaria and the investment opportunities it creates, a look on the sector of Special Investment Purpose Companies/Funds[30], connected with the securitization of real property, including REO acquisitions (*with respect of the individual investment strategy*), or if bank-owned – mainly on such properties. Of the latter, there is only one ranked in Top 10 of Bulgaria's REITs on shareholders' equity in 2012 (*Central Cooperative Bank Real Estate Fund SPIC/REIT*)[31] with the rest of the funds in the hands of other strategic domestic and foreign investors. Overall, the total market capitalization of the special purpose vehicles on the Bulgarian Stock Exchange (BSE) for the fourth quarter of 2012 was 1,715 mln. BGN, representing 17.4% of the total market capitalization of all the companies listed on the Bulgarian Stock Exchange in Sofia[32].

---

[28] Source: citybuild.bg, 2013-06-24 According to Julian Edwards, managing director of Tishman International for Europe. From the International online forum for real estate and investment „RINFOR".

[29] EPRA/IREF, Real Estate in the Real Economy Report

[30] „Special Investment Purpose Joint-Stock Company" (SIPJSC), as stated in the Act on Special Investment Purpose Companies, promulgated by the State Gazette, issue No. 46 dated May 20, 2003. Those are the Bulgarian version of the Real Estate Investment Trusts (REITs)

[31] With 24, 9 mln. BGN in equity, according to the data of Information system for Bulgarian enterprises with rankings based on annual financial statements for the REIT, submitted to BSE or to the FSC prior to 06/20/2013

[32] Compared to the third quarter of 2012 the market capitalization of REITs increased by 0.7% compared to the total market capitalization of companies listed on the BSE. Despite the progressive trend, REITs shares continue

# 4. Managerial Approach on the Subject of Crisis' Heritage

What we discussed so far in terms of crisis' heritage - non-performing loans, foreclosure procedures, eventually real estate owned properties, increased the importance of good corporate governance and risk assessment and management[33] in the financial sector. Many of the financial institutions created subsidiary asset management companies and began a process to effectively manage problem exposures through the acquisition of securities, subsequently transferred to the management companies or sale. The situation also increased the importance to work on early recognition of signs of an increased risk of deterioration in the financial condition of borrowers, as well as on the better management of relationships with these customers[34]. In this regard, there have been positive changes in the organizational structure and behaviour of banks from purely corporate-principal point of view. Minimally, managing OREO require banks to develop a strategy to optimize returns; to assess resource requirements; to strengthen their infrastructure for dealing with OREO and to establish meaningful reporting. They are required to reassess their internal processes, starting from the phase of obtaining, through the period of maintaining and disposition of OREO in terms of adequacy of the policies, practices, procedures, internal controls; to evaluate the validity and quality of all OREO owned; to determine compliance with existing laws and regulations.

For a profitable business venture in the area of OREO management to be extracted out of the country-specifics *(different than the case when bank manages the process on its own)*, a scenario planning method should be imposed as a version of the traditional business analysis. First, there should be an analytical phase, focusing on the most important factors from the environment of OREO property management and from there – to derive assumptions what changes may occur with the environment and how this could affect the company the entrepreneurial initiative. Next, on any possible change with high level of probability, concerning the main factors of the environment, three different scenarios should be formulated in terms of changes in the business model. The process continues with proposals

---

to be undervalued and the shares of nearly all REITs are being traded at a considerably lower price than that of the asset value which makes those shares very attractive to potential investors for specific actions and / or alternative strategies with which the entrepreneur to "meet" the expected changes in the area – both internal and external.

[33] New risk issues also arise, along with OREO management development. Some risk elements are parallel to those found in other banking activities, for example, the potential conflict of interest such as the use of a firm by a banking organization to both provide information on a property's value and list that property for sale on behalf of the banking organization.

[34] Rob Whitmire, senior V.P. of Special Assets Services Group, says, quoted by Bull Realty, that a lender should react swiftly to understand the underlying asset when a loan first shows signs of trouble. The lender should begin compiling reports on the collateral at that time and should "*get all the information from the borrower it can when it still has a good relationship*" with the borrower.

If we look into the business models/scenarios that are possible alternatives for OREO management ventures in Bulgaria, we will see there are three main groups of business models, existing on the market: bank-owned companies (*private and public companies*[35]); joint-ventures between banks and investment funds or/and big real-estate companies; private servicing companies (*from mortgage servicers[36] to property preservation companies, full operational facility management companies or consultants real estate agents, attorneys, appraisers, title companies, etc.*) All of them developed progressively along the boom and bust cycle, connected with the construction and mortgage market. And continue to develop according the country-specifics, no matter if the property is commercial or residential. Because having professionals routinely checking the assets on the grounds ensures that the value doesn't diminish over time and that maintenance costs remain stable, at a reasonable price. Otherwise, it will not be profitable for such assets to be hold in possession. To sell OREO asset in a depressed market is undesirable, but holding it for an extended period, without putting the efforts needed to raise its value, can be a risky and expensive exercise. The analysis for Bulgaria in particular, shows that many banks such as Eurobank, BACB, Raiffeisen, UniCredit Bulbank, Central Cooperative and Central Trade Bank, Pireos Bank, have either already created bank-owned private or public companies (*under the special investment purpose companies regime*) or have established the joint ventures, mentioned above, with big real-estate businesses for the purposes of the management of their property assets, including their lease or sell. Not many independent servicing companies are on the market, but this is where the entrepreneurial opportunities are, as many work within the SPICs allocation of functions regalement[37], managing the properties and leasing them on behalf of the holder *(mainly of business buildings, warehouse, retail and office space),* given the overall state of the market which is to improve.

---

[35] According to the author of the article, the public company, the investment model, is the better approach as it provides more transparency which goes hand-in-hand with the dissemination of good information and enables scrutiny. Scrutiny not only by the shareholders, but also by an „army" of analysts. Scrutiny of the private subsidiary companies is less intense, led by a few multinational and niche investment consultancies, according to the EPRA.

[36] A definition from the „FDIC Law, Regulations, Related Acts" by the Federal Deposit Insurance Corporation goes that a mortgage servicer is a company to which some borrowers pay their mortgage loan payments, if it has bought the mortgage servicing rights from the original mortgage lender.

[37] The legal regulations under the Bulgariam SPIC/REIT's regime, concerning the listed property companies, the so called special purpose vehicles for securitization of real estate, provides for allocation of the functions between servicing companies, depository bank, real estate appraisers, etc.

A prospective entrepreneurial segment with high added value, inside the overall business environment, is the analytical one - preparation of surveys and analyses, forecasts and assessments of the state of the real estate market in the county and development of a tailored strategies for subsequent administration or disposal of those property assets. Though further analysis of banks` business-models (*mentioned above*), reveals that they themselves rely on high level of integration in their OREO management procedures within subsidiary private companies. Why is that? Elementary – it is because of the scales of the market. Bulgaria is a small market and when banks design the management process, they design it in a way that it won't only reduce loses, but will also give them additional profits out of the OREO management specialization. And this can only happen, on a small-scaled market, with the help of a broader and diversified service portfolio, intend to provide services on the overall real estate market. For example – Post banks' OREO management company is active in the following sectors: *real estate advisory, market research, valuations, brokerage, property & facilities management, corporate real estate, real estate asset management, technical services, group properties*, etc. From the subsidiary company's point of view the four main phases of the full asset management cycle should be consolidated. According to the time period and the area of specialization (*with the reporting processes, integrated in each and every other stage*), those phases are: - the evaluation and setup stage (*assignment of qualified, OREO agent in charge, inspection of property conditions' and environmental evaluation; security, safety and compliance issues; maintenance, rehab and marketing plans preparation*); -the marketing stage (*establishment of list/sale price according to predetermined limits; integration into the overall tracking/monitoring system; monthly marketing reports and updates on the strategy and pricing*); -the property/facility management stage (*rehab, maintenance and optimization plans and procedures execution; monthly inspection reports, track and timely payments on property expenses*); -the contract negotiation & closing stage (*negotiation over offers within predetermined authority levels; managing closing process; timely receipt of sale proceeds; billing*).[38]

Though each and every of the stages listed above is oriented towards increasing the overall value of OREO properties and the integral efficiency of the management procedures implemented, for commercial properties, in particular, certain economies on the initial maintenance budget, not only can add value to the property, but can be crucial for the overall economic success of the management process by attracting more tenants and/or higher rental returns in case that a lease strategy is imposed. Cost optimization can be achieved in various ways through overall process optimization as there is no *"one size fits all"* approach, but the greatest potential for savings, concerning commercial property, comes from energy costs. Up to 30% in energy costs savings can be achieved by a facility management company through adjustments of building systems without requiring additional investments thanks to the modern technological solutions (the *so-called building management systems*). This can turn back on its feet a non-viable or even bankrupted commercial property project (*such as a shopping mall*) into a profitable business venture. No need to mention some striking examples of failed construction projects around Europe, mainly in government hands, due to the enormous mainanence costs and bad management for which reorganization and restructuring, including in terms of ownership, is not the best, but the only solution. And its agent is the crisis itself. Such one zombie-project is the new Berlin Brandenburg „ghost"Airport (BER), which according to a recent financial inquiry devours nearly 20 million euro per month, spent on cleaning, security, maintenance, repairs and, above all, energy. And that's just prior to its opening, which was delayed a couple of times already[39]; each and every of this delays accumulating additional expenses, with a need to inject taxpayers' money again and again[40]. So far, airport shareholders - the federal and local governments are backing the loans of the airport operating company, but the more they do it the bigger harm they do on its recreation.

---

[38] According to the full-cycled services portfolio model of the Bankers Asset Management Company, U.S.

## 5. The Behavioural Side of Crisis' Heritage Management

Without being too adverbial, if the analysis is to be full, we should look more closely on a couple of basic strategic thinking and entrepreneurial behaviour principles that should be utilized prior to start tackling properly the financial crisis and harvest the opportunities out of OREO management business ventures. No other reliable scientific method exists when there is no computability of the probability of rare high-impact events (*such as those we are facing today*}, therefore we can only rely on personal preparation in terms of practical skills and applied knowledge acquired. There is a necessity for critical-thinking skills implementation in today's complicated financial and economic environment in order to grab this opportunity.

---

[39] Construction started in September 2006. The airport was originally due to open on Oct. 31, 2011. It was then delayed to June 3, 2012, then to spring 2013, and finally to October 2013 prior to its latest delay as of Jan 2013. Newspapers (Spiegel, http://www.spiegel.de/international/germany/opening-of-berlin-airport-delayed-again-due-to-technical-problems-a-876103.html) claim it may not be able to go into operation until October 2014 or even 2015.

[40] The German government ownes a 26 percent share in the airport; Berlin and Brandenburg own 37 percent each. So far the cost of the project has doubled to €4.3 billion, while latest projections show that hundreds of millions of euros will have to be added to the bill in the form of new construction costs, lost income, possible compensation payments to retailers and airlines for lost revenues, etc. Source: *the above mentioned*.

This process itself, in terms of mastering and learning to implement, is actually quite complicated in its nature, consisting of couple of stages[41].

First of all, it is all about comprehending opportunities' dimensions. Real estate research groups estimate more than four million properties in foreclosure, only in the US, significant part of which are commercial properties. According to Christopher Marinac, Managing Principal & Director of Research, FIG Partners LLC in 2012 U.S. banks had about $10 trillion in total assets of which approximately $300 billion are OREO properties. But that's just the assets that have been foreclosed, without calculating the inventory of potential problems ahead[42] and without calculating the market capitalization of REO properties in the hands of REITs[43].

Following the importance of a comprehensive understanding on the dimensions of the opportunity unveiling, comes the habit of mind, characterized by the comprehensive exploration of issues, ideas and events before accepting an opinion or conclusion. „Exploration of issues" means that the modern-day entrepreneur, REITs' investor, OREO manager, banking or even government official, should be able to extract *evidence* of all relevant contextual factors prior to determining its behaviour as an economic agent. For example, „chronology" is an important contextual factor when dealing with the financial crisis. If the OREO management business opportunity is not (*self*)explained in a chronological way with all the causes-and-effects relationships, then it is hard to predict its further development directions and to react in a profitable way. Not many date back the cause and effect chain that shifted commercial property financing and management form Main Street to Wall Street earlier than 2001, but in fact it was the 1987-1994 period of the last major real estate recession that caused a fundamental restructuring of the values and ownership of commercial real estate (Hunton & Williams, 2011) and marked the beginning of the shift[44].

---

[41] Daskalov. H., Tackling the Debt Crisis Through Utilization of Applied Knowledge, SBS 2012 Edition. Article, nominated for participation in Von Mises Seminar (Genova 2012)

[42] More at Bull Reality Commercial Real Estate Company's blog at:
http://blogs.bullrealty.com/blogs/foreclosure/2012/04/banks-have-more-bad-debt-oreo-properties-coming

[43] North America's listed property sector of 372,7 bln. euro (EPRA index market capitalization) represents 6,7% of the total real estate sector's relative size in North America (5,599.0 bln. euro). In Europe the numbers are respectively 103.1 bln. euro vs. 5,758.6 bln euro in total real estate which is approx. 1/50 ratio (1,8%). This percentage is greatest in the UK (4,7%) , led to only 0,1% in Italy. Source – EPRA, Prudential – Data as of 31 October 2011, Total real estate as of December 2011

[44] On a „*Black Monday*" of October 1987, a stock collapse with an unprecedented size caused „Dow Jones Industrial Average" to fall by 22.6 percent, affected by the savings and loans collapse and ultimately leading to a slump in the United States real estate market. Source: William A. Walsh, Jr., Ian R. D. Labitue and C. Scott Tuthill, Hunton & Williams LLP, How Financial Institutions Can Effectively Manage and Dispose of REO

The proper selection and usage of information to investigate a point of view or a conclusion is another important behavioural aspect in this line of thoughts. Increased accuracy in this aspect allows for greater support and preparation against the unknown while reducing expenses in the long run. It is also qualified as an *'evidence seek'* in the critical-thinking-skill-set in which one should be able to take information from source(*s*) with enough interpretation and evaluation to develop a comprehensive analysis or synthesis. For example, an inappropriate assumption used in determining the fair value of purchased assets may result in a requirement to recognize losses shortly after the inception of the theoretical transaction. Of course, this process can also be backed-up by consultants or executed with the help of specialized software solutions (such as MST applications)[45] and catalyzed by the bigger rate of transparency in market data. It is important though, to remember to question thoroughly even viewpoints of proven experts, no matter their capacity *(as capacity obviously should not matter according to the recent developments in the banking sector),* therefore improving market fundamentals data and performance measurement is the safest way out.

Eventually, the last behavioural aspect in the chain of critical thinking skills, needed for success in the changing financial world, is on the ability to project conclusions and related outcomes *(implications and consequences),* such as those about dealing with future incomes of presently possessed „disturbed" property assets. This can be interpreted as a need to foresee in times of short-sightedness, but that is what this article is all about –uncovering the horizon of opportunities through critical review of the sceptic voices, dominating the financial crisis discussions. As challenging as it may look to be the estimation of the current value of an OREO asset, estimating its future valuation trend requires even greater efforts as it is everything but simple to forecast with certainty in such turbulent times the impact of factors such as the potential new construction in the specific sub-market, the future demand and other essential considerations of importance.

Exactly the lack of focus on the long term-perspective – the risks and opportunities, waiting behind the corner, was in the core of what triggered this chain of events that led to the present state of the financial world. As a result, today's asset property managers – privately hired or by the banks, value more (or at least should) such basic necessities as to plan, but also to assist the positive outcomes; to properly evaluate the technical, physical and financial

---

Assets, Bloomberg Law Reports ─ Banking & Finance, Volume 4, No. 4 (April 2011) and Volume 4, No. 5 (May 2011).

[45] MST (MainStreet Technologies, Inc.) works with senior bank management in financial institutions throughout the U.S. and Asia, implementing application software that addresses the most critical and complex processes in managing loan portfolio risk. The suite of solutions specifically developed for bankers is designed, installed and supported by MST.

characteristics of the residential and commercial properties possessed; to monitor closely the yield and returns over time and to carry out periodically market value reviews, supported by market research for the sectors and locations of interest. In other words – the crisis brought back and increased the value of the classic-style managerial efforts and expertise in areas such as commercial property financing, construction and facility management and gave a good lesson to all the stakeholders in the process not just to hedge the risk, coming out of their irresponsible behaviour, but to go back to the old-fashioned managerial capitalism which is in the base of the crisis heritage management concept.

# 6. Implications & Conclusion

We can conclude by comparing the new business opportunities out of the financial collapse (*which don't only help solve problems but by their entrepreneurial nature create opportunities for employment, service- and process innovations*) to what is presently used in the fight against the insolvency in the banking sector by the national and EU governments. The present measures *(the governmental ones)* are not proactive, and represent bad examples of pro-crisis zombie-economy incentives, instead of good examples of change and opportunity management. Therefore, it can be stated that heritage management begins where zombie-management[46] ends, though those are in fact the two sides of the same coin as zombie-economy also takes its roots in times of financial crisis.

During such times, governments feeling „called" to assist the economy provide bailouts and other „supportive" policies directed towards the financial institutions and private corporations, but instead create zombies that try to preserve for their own sake the benefits that were available during the crisis. An example of a pro-zombie-economy measure or a pro-zombie-economy incentive is the depositors' guarantee. They ought to reflect, but only in the eyes of regulators, the necessity to ease the stress in the banking sector. A deposit insurance scheme, implemented by the government, might definitely sound reasonable, but if the phases of the critical-thinking process, presented above, are comprehended, even on a basic level, it should be clear that the moral hazards as a result/consequence of such actions present by far a bigger threat for the system[47]. Such pro-zombie-measures are nothing but the same measures that brought down the financial sector and were the main cause behind the present „heritage" left to be managed in the finance and real estate sector which only comes to show that Schumpeter's theory on the creative destruction is applicable in this case and proven to be right once again.

---

[46] The overall vicious form of governance of the financial system and the malignant relationship and correlations between zombie-firms, zombie-banks and their governments is defined here as a zombie-management. More on the subject can be found in „Zombie Economy – The Legacy of the Financial Crisis" by Vladimer Papava a well-known economist, author of the „theory of necroeconomics".

A creepy legacy alternative should vanish in order for a vibrant and profitable heritage management opportunity to substitute it. The development of solid OREO management strategies represents an essential part of this opportunity – in terms of internal financial institutions guidance, business models, exercised by private operators, national and pan-European overviews, that would all significantly contribute towards a safer exit out of the banking crisis, representing a chance for new business ventures and new investment opportunities to arise.

Ultimately, in the theory and in practice, comes the phase for evaluation of the possible and reasonable solutions, done in a way which considers with respect the history of the problem, reviews the logic, examines the feasibility and weighs the impact of the solutions proposed, such as those discussed in the paper. Indeed, many are aware that a free and unspoiled market is the best way individual and collective progress can advance, and would passionately argue that banks' business has nothing to do with real-estate management or anything near it, hence everything that originates from this source should be labelled as improper subject of discussion. The present article suggests something else. That the corporate, finance and even public sector representatives can reach significant results in their strive to help the economy recover and grow once again by joining efforts on strategic and operational level for the management of commercial (*but not only*) real estate properties, which have ended up temporary in the hands of lenders, banks, subsidiary companies and REITs, special servicers, OREO property & facility management businesses, etc. Such results can turn out to be far more effective than what has been achieved so far through the implementation of the well-known anti-market bureaucratic policies, proven to be unproductive and in some cases – extremely harmful. Policies which didn't allow neither the banks, nor the markets to function properly and clear themselves form the insolvency burden, which eliminated capitalism from the equation[48].

---

[47] They (such guarantees) might and eventually do lead to reckless behaviour from both depositors in a strive for higher interest rates, regardless of bank's soundness, and from bank's executive officers in strive for bigger market reach with the ultimate bonuses goals.

[48] In his book – „The Black Swan", the author Dr. NassimTaleb says: "*When you remove failure from the economy, you eliminate capitalism.*"An economy that allows failure and goes through failure eliminates

The assets, behind the concept, justifying the need for active professional management of the „heritage", left of the financial collapse, if not else, give chance to many new entrepreneurial and investment opportunities after the long record of unsuccessful financial engineering exercises by the governments that turned the financial institutions into no longer effective capital intermediaries[49]. As if to reinforce the idea that in business the only thing that gives managers a second chance is the crisis itself.

**Acknowledgements**

The author is grateful to Gordon Kerr of Cobden Partners for his helpful comments and suggestions on the subject and also to all the participants in Sofia Business School 2013 Master Classes on Financial and Economic Risk Management where the draft was presented and discussed. Errors and omissions however are entirely the author's responsibility.

**Sources of information**

---

weaknesses and irresponsible behaviors from time to time, strengthening the system – a process also known as antifragility. Therefore, whenever governments take money from taxpayers to bailout "too-big-to-fail"corporate giants (mainly in the finance sector) they do it on the cost of eliminating „failure" as a concept from the free market equation and thus weaken the system even more.

[49] Present days many of the financial institutions that were bailed out with taxpayers money in both Europe and US are even more inefficient at their core function than before. Instead of restructuring and moving back to conservative banking, they provided almost the entire benefit handed out by governments , the ECB or the FED to their own staff. New scandals are emerging everyday concerning fixing libor and euribor and other allegations towards banks' management including cartel in the derivatives market and insider trading.